\documentclass[pre,twocolumn,showkeys,showpacs,superscriptaddress]{revtex4}
\usepackage{graphicx,latexsym,amssymb,amsmath,amsbsy}

\begin{document}

\title{Athermal Jamming vs Thermalized Glassiness in Sheared Frictionless Particles}

\author{Peter Olsson}
\affiliation{Department of Physics, Ume{\aa} University, 901 87 Ume{\aa}, Sweden}
\author{S. Teitel}
\affiliation{Department of Physics and Astronomy, University of
Rochester, Rochester, NY 14627}
\date{\today}

%-------------------------------------------------------------------------
%  ABSTRACT
%-------------------------------------------------------------------------
%\vspace*{1cm}
\begin{abstract}
Numerical simulations of  soft-core frictionless disks in two dimensions are carried out to study behavior of a simple liquid as a function of thermal temperature $T$, packing fraction $\phi$, and uniform applied shear strain rate $\dot\gamma$.  Inferring the hard-core limit from our soft-core results, we find that it depends on the two parameters $\phi$ and $T/\dot\gamma$.   $T/\dot\gamma\to 0$ defines the athermal limit in which a shear driven jamming transition occurs at a well defined $\phi_J$.  $T/\dot\gamma\to\infty$ defines the thermalized limit where an equilibrium glass transition may take place at a $\phi_G$.  This conclusion argues that athermal jamming and equilibrium glassy behavior are not controlled by the same critical point.  Preliminary results suggest $\phi_G <\phi_J$.
\end{abstract}
\pacs{64.70.Q-, 45.70.-n, 64.60.-i}
\maketitle

Seemingly disparate physical systems are found to undergo a transition from a flowing liquid-like state to a rigid but disordered solid state upon varying some external control parameter \cite{Jaeger}.  Athermal granular particles undergo a sharp jamming transition as the packing fraction $\phi$ is increased above a critical value $\phi_J$.  Simple liquids may freeze into a disordered glass as temperature $T$ is decreased below the glass transition $T_g$.  Foams may cease flowing and show elastic response as the applied shear stress $\sigma$ is decreased below the yield stress $\sigma_Y$.  An interesting conjecture by Liu and Nagel and co-workers \cite{LiuNagel,OHern} attempted to unite these phenomena into a single {\it jamming phase diagram}, with the three axes $\phi$, $T$, and $\sigma$.  

In the equilibrium $\phi-T$ plane, it may be expected that as $\phi$ decreases, the glass transition $T_g(\phi)$ decreases and vanishes at a particular value $\phi_G$.  In their original jamming phase diagram, Liu and Nagel proposed that this $\phi_G$ was the same value as the $\phi_J$ that locates the athermal jamming transition. The point $T=0$, $\phi=\phi_J$, $\sigma=0$ thus locates a special critical {\it point J} that controls not only athermal jamming, but also finite temperature glassy behavior.  If correct, such a scenario could offer a new approach towards understanding equilibrium glassy behavior, by studying the effect of thermal fluctuations about the presumably simpler and better understood jamming point J. 

However, for protocols restricted to the $\phi-T$ plane, it is now generally accepted from theoretical mean-field calculations \cite{Krzakala,Mari,Parisi,Jacquin} and demonstrated by numerical simulations \cite{Chaudhuri,Vagberg,Schreck} that the location of the $T=0$ jamming transition is in general protocol dependent. A continuous range of values for $\phi_J$ may be found, depending upon the particular ensemble of initial states, as well as on the rates of compression or cooling.  Rapid compression or cooling from random configurations leads to a lower bound, often associated with {\it random close packing}. It remains unclear if a well defined nontrivial upper bound for $\phi_J$ exists for vanishingly slow compression or cooling, or if crystallization must in principle occur for sufficiently slow rates \cite{Donev}.  In contrast, a true equilibrium glass transition $T_g$ should be protocol independent.   The connection between athermal jamming and an equilibrium glass transition is therefore subtle, if one stays in the $\phi-T$ plane.

Here we address this problem by looking along the third axis of the Liu-Nagel phase diagram, considering the behavior of systems undergoing uniform steady state shear, with a fixed applied shear strain rate $\dot\gamma$.  
Considering a simple model of overdamped, frictionless disks in two dimensions (2D), we have earlier shown \cite{Vagberg} that athermal steady state shearing  defines a statistical ensemble of states, such that there is a uniquely defined shear-driven jamming transition $\phi_J$ in the limit of vanishingly small strain rate, $\dot\gamma\to 0$, independent of the initial configuration.  We now extend these investigations to finite $T$. We argue that the distinction between shear driven athermal jamming and thermalized glassy behavior may be considered within a strictly hard-core model.  Using simple dimensional arguments, and by inferring the hard-core limit from simulations of soft-core particles, we show that athermal jamming and thermal glassy behavior are described by opposite limits of a key control parameter of the hard-core system.  We thus conclude there is no reason to expect that these two phenomena are controlled by the same physical processes, nor any reason to expect $\phi_G=\phi_J$.

Our model is one that has been extensively studied at $T=0$ \cite{OHern}.  We use a bidisperse mixture of frictionless soft-core disks in 2D, with equal numbers of large and small particles with diameter ratio $d_l/d_s=1.4$. The soft-core interaction between overlapping particles $i$ and $j$ is harmonic, $V(r_{ij})\equiv\epsilon\tilde V(r_{ij})= \frac{1}{2}\epsilon\delta_{ij}^2$, where $\delta_{ij}=(1-r_{ij}/d_{ij})$ is the relative particle overlap, $r_{ij}$ is the particles center to center distance and $d_{ij}$ is the sum of their radii.  We use Durian's bubble dynamics \cite{Durian} of overdamped particles with a viscous dissipation with respect to an imposed average linear shear velocity flow in the $\hat {\bf x}$ direction,
\begin{equation}
\lambda\left[\frac{d{\bf r}_i}{dt}-y_i\dot\gamma\hat {\bf x}\right]=-\sum_j\frac{dV(r_{ij})}{d{\bf r}_i}+{\boldsymbol\zeta}_i\enspace,
\label{eDurian}
\end{equation}
where $\lambda$ is the viscous damping coefficient.  Temperature is modeled by a random Langevin thermal force ${\boldsymbol\zeta}_i$ satisfying,
\begin{equation}
\langle{\boldsymbol\zeta}_i\rangle=0\enspace, \quad\langle {\boldsymbol\zeta}_i(t){\boldsymbol\zeta}_j(t^\prime)\rangle=2\lambda T\delta_{ij}\delta(t-t^\prime){\bf I}\enspace, 
\end{equation}
with ${\bf I}$ the identity tensor.  We use Lees-Edwards boundary conditions \cite{LeesEdwards} to apply a uniform shear strain rate $\dot\gamma$ to a box of length and height $L$ containing $N$ particles.  

Defining the time constant $t_0=\lambda d_s^2/T$, we can cast the above equation of motion into dimensionless form by dividing each term by $T/d_s$ to get,
\begin{equation}
\dfrac{d\tilde{{\bf  r}}_i}{d\tilde t}-\tilde y_i(\dot\gamma t_0)\hat {\bf x}=-\dfrac{\epsilon}{T}\sum_j\dfrac{d\tilde V}{d\tilde{{\bf r}}_i}+\tilde{\boldsymbol\zeta}_i\enspace,
\label{ehc}
\end{equation}
with dimensionless variables $\tilde{\bf r}_i\equiv {\bf r}_i/d_s$, $\tilde t\equiv t/t_0$, and noise $\tilde{\boldsymbol\zeta}_i$ satisfying the correlation $\langle 
\tilde{\boldsymbol\zeta}_i(\tilde t)\tilde{\boldsymbol\zeta}_j(\tilde t^\prime)\rangle =2\delta_{ij}\delta(\tilde t-\tilde t^\prime){\bf I}$.  
In the hard-core limit, $\epsilon/T\to\infty$, the first term on the right hand side acts to provide an excluded volume effect, preventing particle overlaps but introducing no energy or time scale.  Behavior in this sheared hard-core limit is thus entirely determined by two dimensionless parameters: the packing fraction $\phi$, and the P{\'e}clet number $\dot\gamma t_0\propto \dot\gamma/T$ 
(for later comparisons, we will find it convenient to phrase our discussion in terms of the inverse P{\'e}clet number $\propto T/\dot\gamma$).
This immediately yields one of our main conclusions.  The effects of temperature and shear on the hard-core system enter only via the combination $T/\dot\gamma$.  The limit in which athermal jamming takes place corresponds to $T\to 0$ first, followed by $\dot\gamma\to 0$, i.e. the limit $T/\dot\gamma\to 0$.  The limit in which thermalized glassy behavior takes place corresponds to $\dot\gamma\to 0$ first, followed by $T\to 0$, i.e. the limit $T/\dot\gamma\to\infty$.  As athermal jamming and thermalized glassy behavior occur at the extreme opposite limits of the control parameter $T/\dot\gamma$, there is no reason to expect that jamming and the glass transition share a common physical mechanism, or that they occur at the same value of the packing fraction $\phi$.

We now consider the shear viscosity $\eta\equiv \sigma/\dot\gamma$.  In terms of the dimensionless variables defined above, the dimensionless viscosity $\tilde\eta$, with a well-defined hard-core limit, is $\tilde\eta=\eta/\lambda d_s^2$.  By measuring the viscosity $\eta(\phi,\dot\gamma, T)$ of soft-core particles, we will infer the hard-core limit $\tilde\eta_{\rm hc}(\phi, T/\dot\gamma)$.  The athermal jamming transition is then defined by the $\phi_J$ where $\tilde\eta_{\rm hc}(\phi_J, 0)\to\infty$, while the hard-core glass transition is defined by the $\phi_G$ where $\tilde\eta_{\rm hc}(\phi_G,\infty)\to\infty$.

Our simulations are at a fixed packing fraction $\phi$, using $N= 65536$ particles  so that finite size effects are negligible for the parameter ranges utilized in this work.  The elastic part of the stress tensor $p_{\alpha\beta}$ is computed from the contact forces in the usual way \cite{OHern}; the elastic part of the pressure is $p\equiv (p_{xx}+p_{yy})/2$ and the shear stress is $\sigma=p_{xy}$. Henceforth we measure length in units such that $d_s=1$, energy in units such that $\epsilon=1$, and time in units such that $\lambda d_s^2/\epsilon =1$.  In these units we have $\tilde\eta = \eta$.  Similar shear driven simulations at finite $T$ have been carried out for {\it underdamped} particles by others \cite{Haxton1,Haxton2,Otsuki}.

\begin{figure}[h!]
\begin{center}
\includegraphics[width=3.4in]{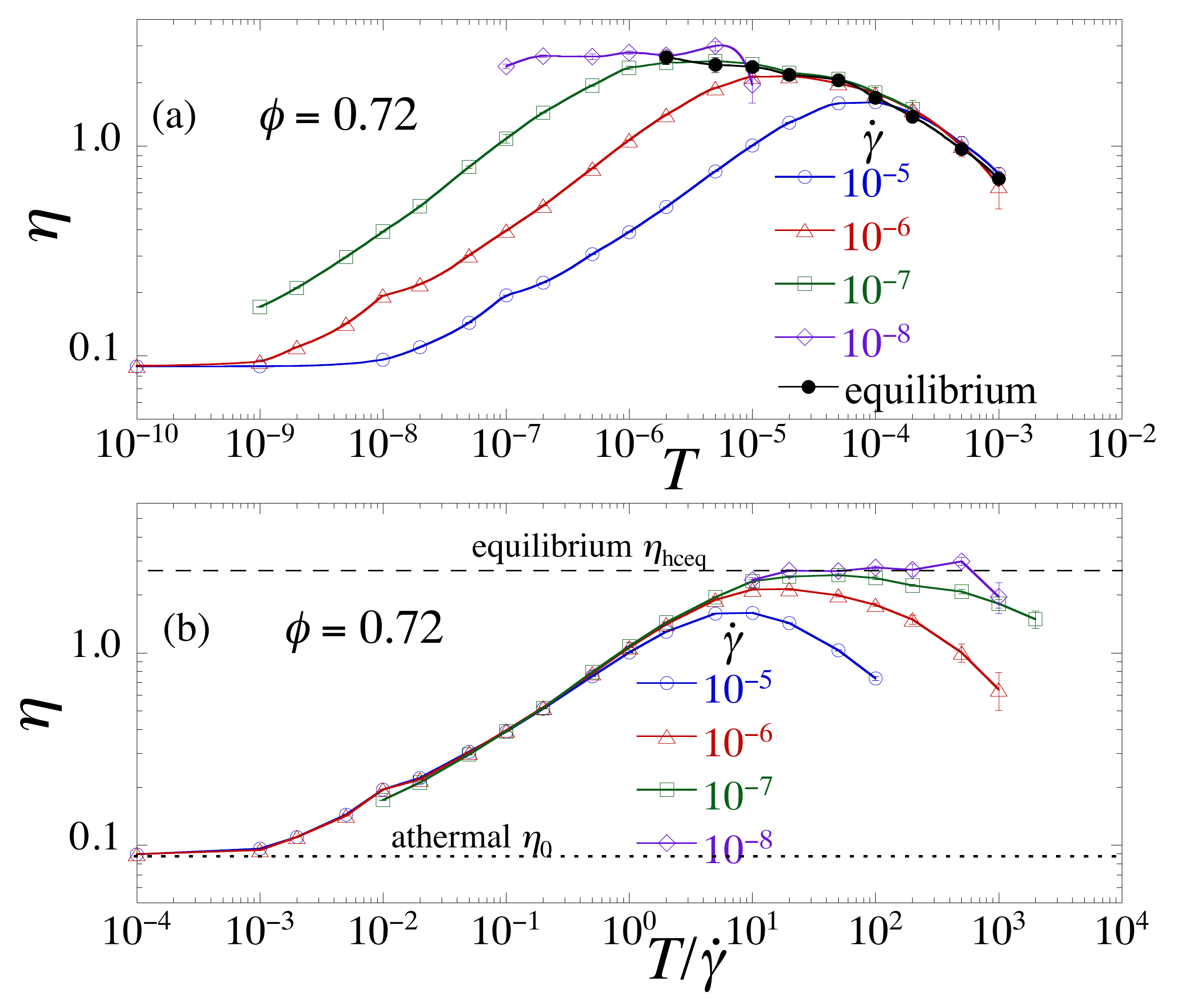}
\caption{Shear viscosity $\eta$ vs (a) $T$ and (b) $T/\dot\gamma$, at fixed packing fraction $\phi=0.72$ below jamming, for several different shear strain rates $\dot\gamma$.  In (a) solid dots are the linear response $\eta_{\rm GK}$ computed from the Green-Kubo formula in equilibrium.  In (b) the horizontal dashed line is the hard-core equilibrium limit $\eta_{\rm hceq}$; the horizontal dotted line is the athermal value $\eta_0$ at $T=0$.
}
\label{f1}
\end{center}
\end{figure}

We consider first a moderately dense value of $\phi=0.72$, but still well below the jamming $\phi_J\approx 0.843$ \cite{OlssonTeitelPRE}.  
In Fig.~\ref{f1}a we plot our results for $\eta$ vs $T$ for several different values of $\dot\gamma$.  We also show the linear response $\eta_{\rm GK}$, computed using the Green-Kubo formula \cite{LeesEdwards} applied to equilibrium ($\dot\gamma=0$) simulations of Eq.~(\ref{eDurian}) \cite{note}.  We see that at high $T$ all the data collapse to a common curve; $\eta$ is independent of $\dot\gamma$ indicating a linear rheology where thermal fluctuations dominate over shear induced fluctuations. 
As $T$ decreases, $\eta_{\rm GK}$ increases and appears to saturate to a fixed value, which is just the hard-core value of viscosity in thermal equilibrium, $\eta_{\rm hceq}$.  As $T$ decreases at finite $\dot\gamma$, however, $\eta$ increases to a peak value $\eta_{\rm peak}(\phi, \dot\gamma)$ at a temperature $T_{\rm peak}(\phi,\dot\gamma)$, and then decreases to a finite value as $T\to 0$.  
As $\dot\gamma$ decreases, $\eta_{\rm peak}$ saturates to $\eta_{\rm hceq}$.  

On the low $T$ side of $\eta_{\rm peak}$, the rheology is highly nonlinear ($\eta$ varying with $\dot\gamma$) until converging to a common value as $T\to 0$.  This $T=0$ value is just the athermal viscosity $\eta_0$.
This low $T$ side of $\eta_{\rm peak}$ can be better understood by plotting $\eta$ vs $T/\dot\gamma$, as shown in Fig.~\ref{f1}b. We now see that the data below $\eta_{\rm peak}$ for different $\dot\gamma$ collapse to a common curve.  This common curve, as $\dot\gamma\to 0$, represents the hard-core limit predicted by Eq.~(\ref{ehc}), which we see is a smooth sigmoidal shaped curve increasing monotonically from $\eta_0$ in the athermal $T/\dot\gamma\to 0$ limit, to $\eta_{\rm hceq}$ in the thermalized $T/\dot\gamma\to\infty$ limit. Data at finite $\dot\gamma$ to the right of $\eta_{\rm peak}(\dot\gamma)$, that fall below this limiting $\dot\gamma\to 0$ curve, represent the parameter region where soft-core effects are important.  From Fig.~\ref{f1}a we conclude that as $\dot\gamma\to 0$, the hard-core region of the system gets pushed down to $T\to 0$.  The thermalized limit, $\lim_{T\to0}[\lim_{\dot\gamma\to0}\eta] = \eta_{\rm hceq}$, becomes singularly decoupled from the athermal limit, $\lim_{\dot\gamma\to 0}[\lim_{T\to0}\eta]=\eta_0$.

The non-monotonic behavior of $\eta(T)$ at fixed $\dot\gamma$ can be understood as due to competing effects of thermal fluctuations on the hard-core vs the soft-core regions of the system behavior. As $T$ increases in the hard-core region, the forces associated with collisions increase and hence pressure $p$ increases.  Since particles cannot pass through each other, it is difficult for shear stress to relax and so as $p$ increases, so does $\sigma$ and hence $\eta$.  As $T$ increases further, however, one enters the soft-core region where there is enough thermal energy for particles to press into each other.  Particles may now squeeze past each other, allowing for more rapid relaxation of shear stress, with a decrease in $\sigma$ and hence $\eta$.

\begin{figure}[h!]
\begin{center}
\includegraphics[width=3.4in]{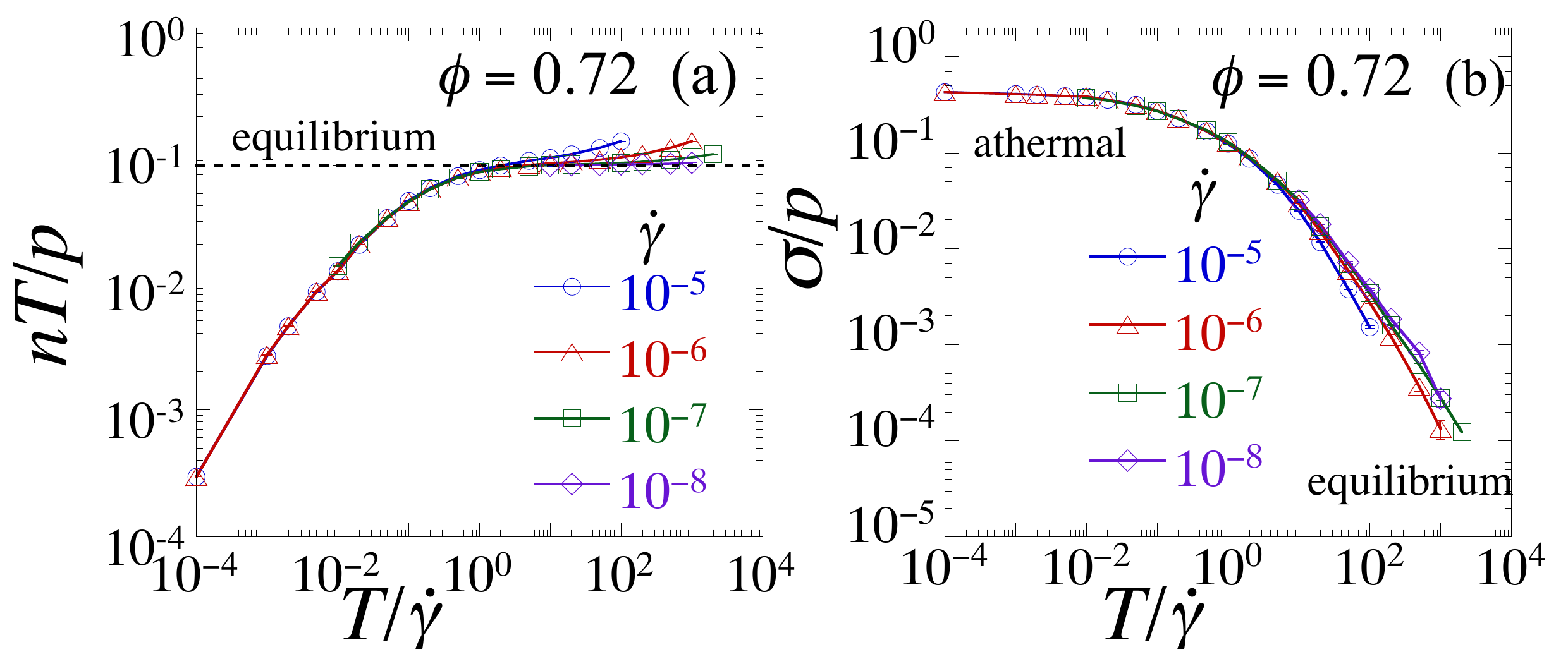}
\caption{a) Inverse reduced pressure $nT/p$ and b) stress anisotropy $\sigma/p$ vs  $T/\dot\gamma$, at packing fraction $\phi=0.72$ for several different shear strain rates $\dot\gamma$.  Dashed horizontal line in (a) is the hard-core equilibrium value of $nT/p$.}
\label{f2}
\end{center}
\end{figure}

It is interesting to consider how other quantities depend on the hard-core parameter $T/\dot\gamma$.  In Fig.~\ref{f2}a we plot the inverse reduced pressure $nT/p$, with $n\equiv N/L^2$ the density of particles, vs $T/\dot\gamma$. In the thermalized limit $T/\dot\gamma\to\infty$ we see that the curves, as $\dot\gamma\to 0$, approach the hard-core equilibrium value, as we have computed independently from pair correlations \cite{Luding} evaluated in equilibrium hard-core Monte Carlo simulations.  In the athermal limit $T/\dot\gamma\to 0$, $nT/p\to 0$ as is expected.  In Fig.~\ref{f2}b we plot the stress anisotropy $\sigma/p$ vs $T/\dot\gamma$.  In the thermalized $T/\dot\gamma\to\infty$ limit we find $\sigma/p\to0$ as expected; as $\dot\gamma\to 0$, the thermalized system maintains a finite pressure but no shear stress.  In the athermal $T/\dot\gamma\to0$ limit, $\sigma/p$ approaches a finite value, as has been observed earlier \cite{Roux}; in the athermal limit, both $\sigma$ and $p$ vanish as $\dot\gamma\to 0$, but do so in a way that their ratio becomes constant. This appearance of anisotropy as one moves from the thermalize to the athermal limit has recently been observed in experiments on colloidal particles \cite{Schall}.

\begin{figure}[h!]
\begin{center}
\includegraphics[width=3.4in]{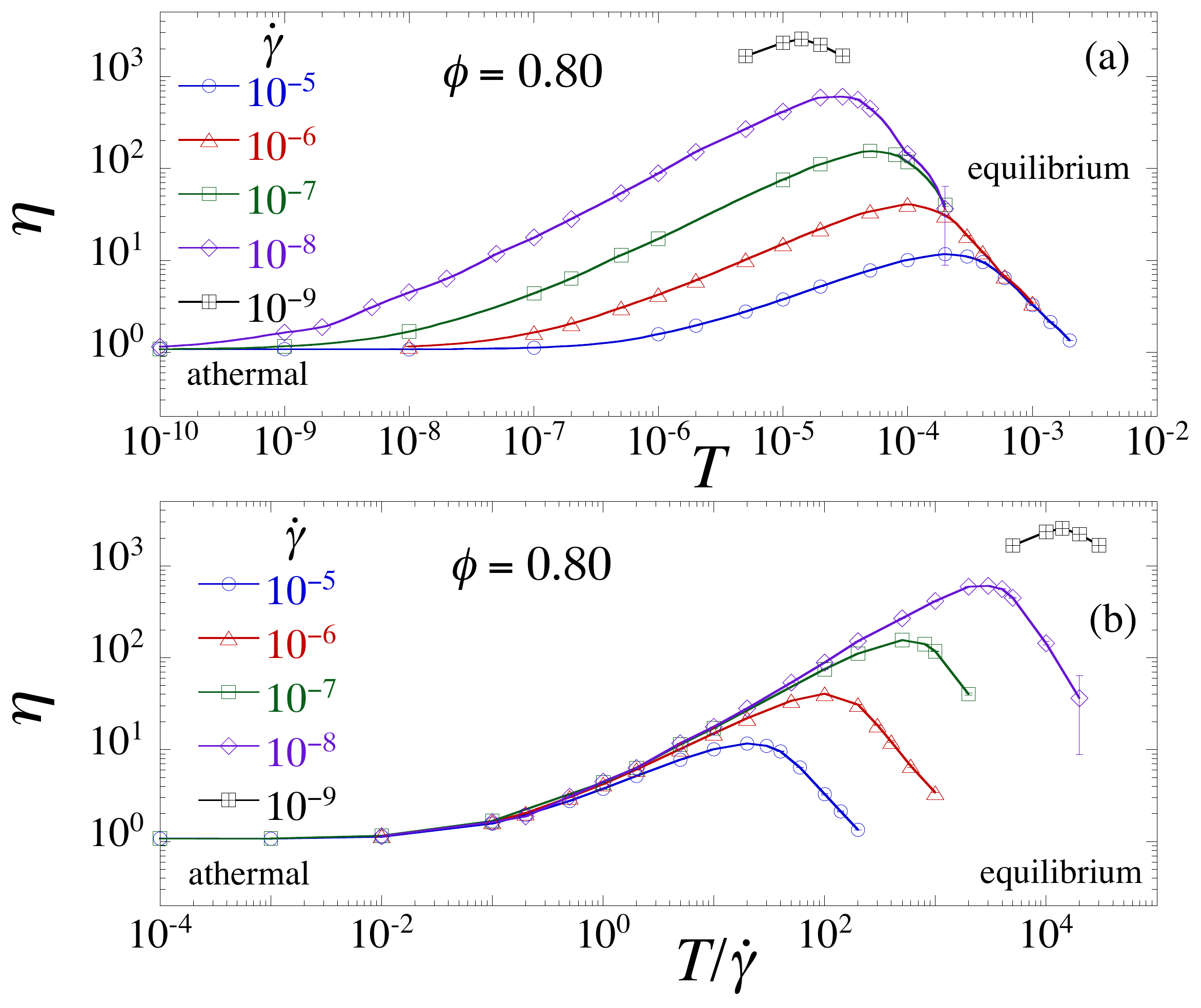}
\caption{Shear viscosity $\eta$ vs (a) $T$ and (b) $T/\dot\gamma$, at fixed packing fraction $\phi=0.80$ closer to the jamming $\phi_J=0.843$, and several different shear strain rates $\dot\gamma$.  
}
\label{f3}
\end{center}
\end{figure}

Returning to viscosity $\eta$, we now consider behavior at higher packing fractions $\phi$.
In Fig.~\ref{f3} we show results for $\eta$ for different values of $\dot\gamma$ at the higher value of $\phi=0.80$.  As in Fig.~\ref{f1}, we see that the data collapses on the high $T$ side of $\eta_{\rm peak}$ when plotted vs $T$, but collapses on the low $T$ side of $\eta_{\rm peak}$ when plotted vs $T/\dot\gamma$.  Unlike Fig.~\ref{f1} however, we see that $\eta_{\rm peak}$ continues to increase as $\dot\gamma$ decreases, with no sign yet of saturating.  Our data in this high $T/\dot\gamma$ limit is not at sufficiently small $\dot\gamma$ to have reached the hard-core limit.  Equilibrium simulations at this high $\phi$ cannot be sufficiently equilibrated to directly compute $\eta_{\rm hceq}$.

\begin{figure}[h!]
\begin{center}
\includegraphics[width=3.4in]{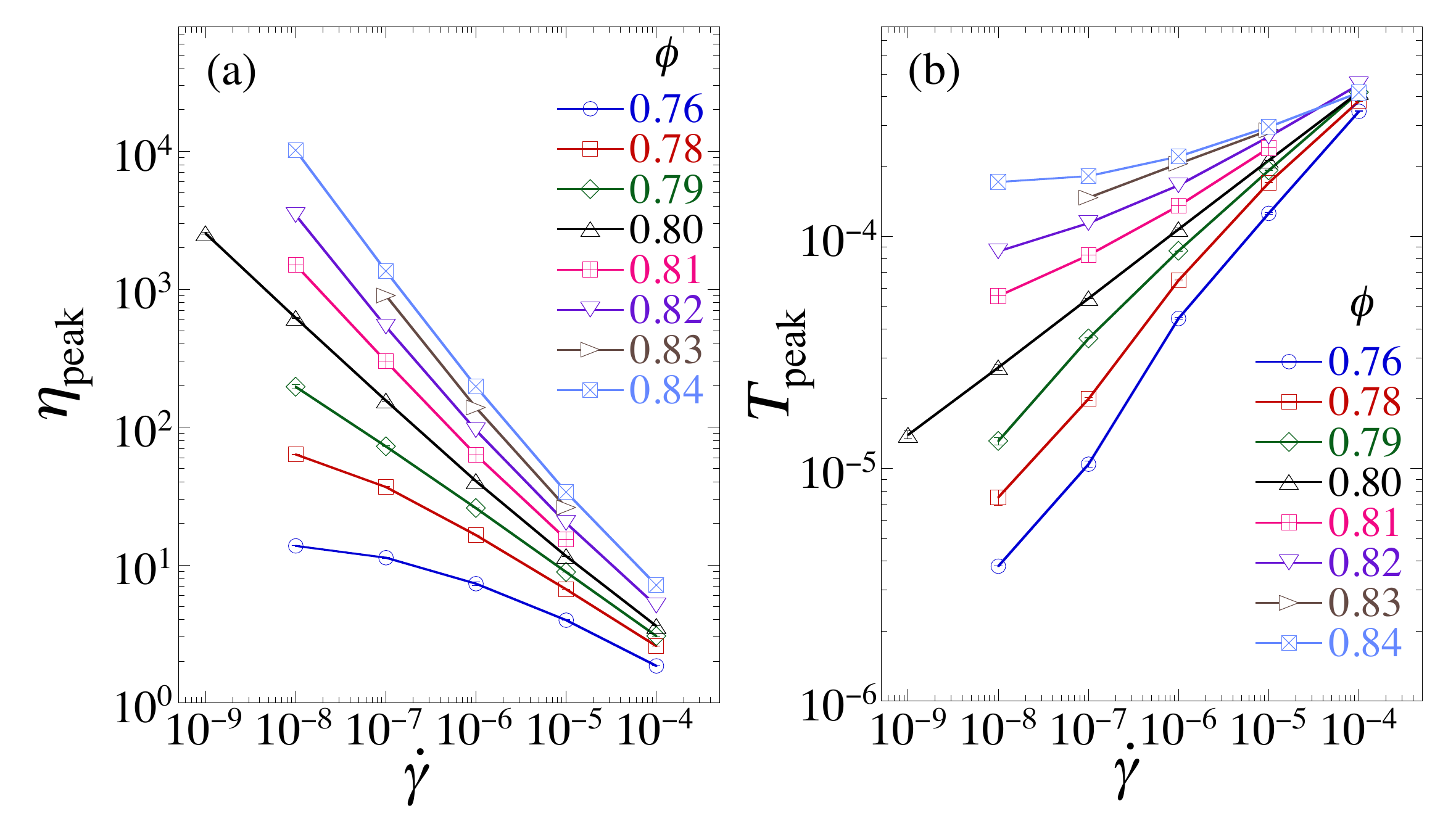}
\caption{a) Peak value of viscosity $\eta_{\rm peak}$, and (b) location of the peak viscosity $T_{\rm peak}$, vs strain rate $\dot\gamma$ for different packing fractions $\phi$.
}
\label{f4}
\end{center}
\end{figure}

That the hard-core  viscosity $\eta_{\rm hc}(\phi,T/\dot\gamma)$ appears to be monotonically increasing as $T/\dot\gamma$ increases, and that, comparing Figs.~\ref{f1} and \ref{f3}, $\eta_{\rm hceq}$ appears to be increasing much more rapidly than $\eta_0$ as $\phi$ increases, suggests a $\phi_G < \phi_J=0.843$, in agreement with recent equilibrium simulations of soft-core particles in three dimensions \cite{BerthierWitten,Xu}.  In Fig.~\ref{f4}a we plot $\eta_{\rm peak}$ vs $\dot\gamma$ for several different values of $\phi$, while in Fig.\ref{f4}b we plot $T_{\rm peak}$.
Since, as $\dot\gamma\to 0$,  $\eta_{\rm peak}\to\eta_{\rm hceq}$,   $\eta_{\rm peak}$ should stay finite  for $\phi<\phi_G$, while $\eta_{\rm peak}\to\infty$ for $\phi\ge\phi_G$. Looking at the raw data in Fig.~\ref{f4}a, such a change in behavior appears to happen at $\phi^*\approx 0.80$.  $T_{\rm peak}$ shows a similar marked change in behavior at the same $\phi^*$, with $T_{\rm peak}\to 0$ as $\dot\gamma\to 0$ for $\phi\le \phi^*$, while $T_{\rm peak}$ is decreasing much more slowly, and perhaps saturating to a finite value, for $\phi>\phi^*$.

\begin{figure}[h!]
\begin{center}
\includegraphics[width=3.4in]{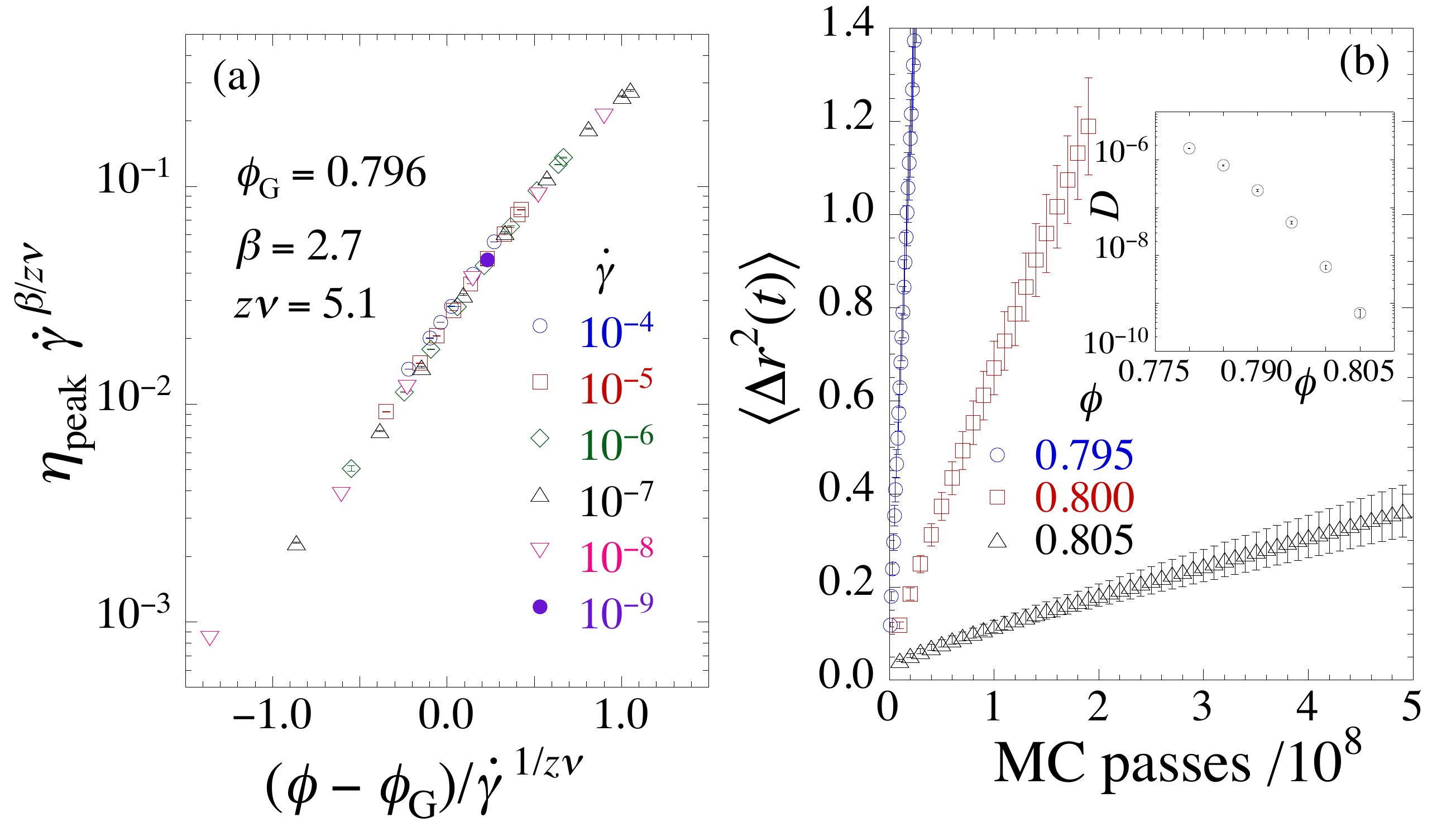}
\caption{a) 
(a) Scaling collapse of $\eta_{\rm peak}$ according to the scaling relation of Eq.~(\ref{escale}).
(b) Mean squared displacement $\langle \Delta r^2\rangle$ vs MC passes for different $\phi$. Inset shows diffusion constant $D$ vs $\phi$.
}
\label{f5}
\end{center}
\end{figure}

We can try to quantify this behavior with a critical scaling analysis.  Assuming the usual algebraic scaling of a continuous critical point at $\phi_G$, we expect $\eta$ to satisfy a scaling relation,
\begin{equation}
\eta(\phi,T,\dot\gamma)=b^{\beta/\nu}f(\delta\phi b^{1/\nu},\dot\gamma b^z, Tb^w)
\end{equation}
where $\delta\phi\equiv\phi-\phi_G$ and $b$ is an arbitrary length rescaling factor.  The peak in $\eta$ as $T$ varies for fixed $\phi$ and $\dot\gamma$ will occur when the scaling function $f(x,y,z)$ has a maximum at some $z_{\rm peak}(x,y)$.  This leads to the scaling equation,
\begin{equation}
\eta_{\rm peak}(\phi,\dot\gamma)=b^{\beta/\nu}g(\delta\phi b^{1/\nu}, \dot\gamma b^z)
\label{escale0}
\end{equation}
with $g(x,y)\equiv f(x,y,z_{\rm peak}(x,y))$.
Choosing $b=\dot\gamma^{-1/z}$ then gives
\begin{equation}
\eta_{\rm peak}\dot\gamma^{\beta/z\nu}=h(\delta\phi/\dot\gamma^{1/z\nu})
\label{escale}
\end{equation}
with $h(x)\equiv g(x,1)$.  Expanding $h(x)$ as a polynomial in $x$, we fit our data in Fig.~\ref{f4}a to the scaling form of Eq.~(\ref{escale}), keeping $\beta$, $z\nu$, $\phi_G$, and the polynomial coefficients as free fitting parameters.  Restricting to data with $\dot\gamma\le 10^{-6}$ our fit yields the data collapse shown in Fig.~\ref{f5}a, with fitted values $\phi_G\approx0.796$, $\beta\approx2.7$, $z\nu\approx5.1$.
Choosing $b=|\delta\phi|^{-\nu}$ in Eq.~(\ref{escale0}) to get $\eta_{\rm peak}=|\delta\phi|^{-\beta}g(\pm 1, \dot\gamma/|\delta\phi|^{z\nu})$, we see that $\beta$ is just the exponent that describes the algebraic divergence of the thermalized hard-core viscosity as $\phi\to\phi_G$.

Although our fit in Fig.~\ref{f5}a appears to look good, we find that the values of our fitted parameters are somewhat sensitive to the range of data that we use.  As we restrict data to a smaller window of $\dot\gamma<\dot\gamma_{\rm max}$ the fitted value of $\phi_G$ appears to increase slightly, while the $\chi^2/$dof decreases.  For our smallest window with $\dot\gamma_{\rm max}=10^{-6}$ we find $\chi^2/$dof $\simeq 6$, suggesting a fit that is still less than ideal.  Such behavior could be due to the effect of corrections to scaling, as we have previously shown to be important at the athermal jamming transition \cite{OlssonTeitelPRE}. 

As an alternative approach to locating the thermalized $\phi_G$, we have carried out independent equilibrium ($\dot\gamma=0$) Monte Carlo (MC) simulations of hard-core disks.  At each step of the simulation a particle is picked at random and displaced a distance $\delta{\bf r}$ with $\delta x, \delta y$ chosen from a uniform distribution on the interval $(-0.05, 0.05)$.  If no particle overlap results, the move is accepted, otherwise it is rejected.  $N$ such steps constitutes one MC pass and represents one unit of time. With this MC dynamics we compute the single particle mean square displacement, $\langle \Delta r^2(t)\rangle \equiv (1/N)\sum_{i=1}^N\langle|{\bf r}_i(t)-{\bf r}_i(0)|^2\rangle$.  In Fig.~\ref{f5}b we show our results for several different $\phi$.  We see that $\langle\Delta r^2\rangle$ continues increasing with time for $\phi\lesssim 0.805$, suggesting a finite diffusion constant $D$, however $D$ is rapidly decreasing as $\phi$ increases (see inset to Fig.~\ref{f5}b).  These results thus suggest a $\phi_G\gtrsim 0.805$.

The discrepancy in the estimate of $\phi_G$ from our viscosity vs our diffusion data could be due to several factors: (i) the neglect of corrections to scaling in our analysis of $\eta_{\rm peak}$; (ii) it may be that the scaling shown in Fig.~\ref{f5}a reflects a mode coupling transition \cite{modeCoupling,Brambilla} that gets cut off by some other physical mechanism on longer time scales as $\dot\gamma$ decreases; (iii) diffusion at the higher $\phi$ seems correlated with increased particle segregation \cite{OlssonTeitelPRE}, and so may be reflecting an approach to a true phase separated equilibrium \cite{Donev} rather than a metastable glassy state.
We note that recent shearing simulations in three dimensions  \cite{Ikeda} similarly suggest a lower value for $\phi_G$ from viscosity measurements than was previously found from relaxation time measurements of both hard-core \cite{Brambilla} and soft-core \cite{BerthierWitten} particles in equilibrium.  The precise value of $\phi_G$ and a complete understanding of the thermal glass transition thus remains for future work.

To conclude, we have demonstrated that in the hard-core limit, a system of overdamped steady-state sheared particles is characterized not only by the packing fraction $\phi$, but by an additional control parameter $T/\dot\gamma$.  The limit $T/\dot\gamma\to 0$ corresponds to the athermal limit in which a sharp jamming transition at $\phi_J$ is clearly observed.  The limit $T/\dot\gamma\to\infty$ corresponds to the thermalized limit where equilibrium glassy behavior is observed. That the athermal and the thermalized regions are at opposite limiting values of $T/\dot\gamma$ argues that there is no reason to expect athermal jamming and thermalized glassy behavior to be controlled by the same critical point.  Our results find behavior consistent with a $\phi_G<\phi_J$, however we can not yet be conclusive about the exact value of $\phi_G$.  As we were finishing this work, we learned of recent work by Ikeda et al. \cite{Ikeda} who have reached similar conclusions.

We thank T. Haxton, A. J. Liu, and P. Schall for helpful discussions.  This work was supported by NSF grant DMR-1205800 and Swedish Research Council grant 2010-3725. Simulations were performed on resources
provided by the Swedish National Infrastructure for Computing (SNIC) at PDC and HPC2N.


\begin{thebibliography}{99}
\bibitem{note}For these equilibrium simulations we use only $N=64$ particles, and even so we have found it exceedingly difficult to well equilibrate at the lowest temperatures.
\end{thebibliography}
\end{document}